# In the Red(dit): Social Media and Stock Prices

James Baker

Rutgers University, Department of Economics

April 30, 2021

Thesis Advisor: Dr. Eugene White

\

Abstract: Spearheaded by retail traders on the website reddit, the GameStop short squeeze of early 2021 shows that social media embeds information that correlates with market movements. This paper seeks to examine this relationship by using daily frequencies of classified comments and buzzwords as additional factors in a Fama-French three factor model. Comments are classified using an unsupervised clustering method, while past studies have used pretrained models that are not specific to the domains being studied.

1. **Acknowledgements**

I would like to thank my advisor, Dr. White, for providing an overwhelming amount of guidance, patience and support during this project. I would also like to thank Dr. Sigman for her assistance in her role as teaching the Honors Thesis seminar. I would also like to thank all of my peers also working on Honors Theses, and congratulate them for all of their hard work. Two of my peers, Madhushalini Pillalamarri and Audrey Somalwar, had the same advisor as me, and I would like to give them special thanks for their assistance.

2. **Introduction**

This paper was written as a two-semester honors thesis project during the Fall 2020 to Spring 2021. I spent most of the summer sifting through topics, including patents, blood diamonds and police brutality. None of them really stuck. However, on September 16, 2021 I sent Dr. White this email:

> ” In a shocking turn of events, I have found another thing I would like to research. I would like to see if Twitter "coverage" of a publicly traded stock or related phrase ("google" and "search engine") can predict the daily returns of the stock, or changes in the highs/lows/volume of trades. The theory here being that investors' valuation of a stock may be reinforced or informed based on their perception of how others think about the companies (sort of like herd behavior) or their familiarity with the firm in general. If there is an effect, I would particularly like to examine whether this effect has gotten stronger during corona times, as many journalists are claiming that now that everyone is sitting at home with their commission-free trading apps like RobinHood, there are tons of amateur investors in the scene, who may be prone to "window shopping" for hot stocks that make the news. Essentially, the question is "did this picture https://twitter.com/elonmusk/status/150387544402575360 affect the stock market?".

Dr. White was initially skeptical that any social media data could have any impact on the markets. It seemed unlikely that the average Twitter, Reddit or Pinterest user knew any more about the behavior of investors or market conditions than Wall Street analysts. Nonetheless, I chose this as my topic. Initially, I planned to use Twitter data, but I did not have the funds to pay for the massive volume of tweets I would need. So, I limited my scope to only reddit. There were a few financial subreddits, but I chose one known as “/r/wallstreetbets”, as it had a reputation for having a very unique community with its own slang (diamond hands, tendies and stonks) and rambunctious subculture, which I found interesting. All of this transpired before January 2021, when suddenly all anyone wanted to talk about was Gamestop, Reddit and Robinhood.

To put it briefly, a hedge fund known as Melvin Capital had shorted the stock of Gamestop, a retailer of video games and video game consoles. At some point, small retail investors, many of them using a trading platform known as Robinhood, decided to thwart the hedge fund and invest in Gamestop. Contributors to wallstreetbets were among the most enthusiastic proponents of this. Thus, in this way, the comments on the subreddit can be seen as providing information as to the herd behavior of investors. This is not necessarily the same as providing information about fundamental or technical factors, but is a clear example of information moving  asset prices embedded in social media. Eventually, the situation made the news and there were even Congressional hearings, wherein the CEO of reddit claimed that the investment advice on his website was "probably among the best" (NYtimes 2019).

The finance-bro boys club that is "/r/wallstreetbets" is not unprecedented. Financiers, traders and the like have always liked to mingle and form communities. For example, the fraternity Kappa Phi Beta's exclusive Wall Street chapter, which was established in 1929 (Wall Street Journal 2014). Every year, it admits only a handful of members, and its current roster includes titans of finance like Michael Bloomberg, Founder of his namesake corporation, Jon Corzine, former CEO of Goldman Sachs, and Laurence Fink, CEO of Blackrock. On the other hand, the website Reddit was established in 2005. Each day, roughly 50 million people use Reddit, and its users may consist of more than a fifth of Americans between 28 and 29 (Oberlo 2019).

Despite some differences in accessibility, both elite old boys networks and modern social media websites are communities where members may learn from, interact with and emulate the behavior of each other. Just as the discussions between members of Kappa Phi Beta may yield them valuable financial information and insights, Twitter tweets, Facebook posts and LinkedIn updates may be sources of information as to how the market will move. Alternatively, these communities may encourage bubbles. An investor has to make a decision to engage in "noise trading" (Black 1986), such as bidding up the price of GameStop to $300. They can be induced to do this via looking at past returns of such strategies (DeLong, Shleifer, Summers, and Waldmann, 1990), or by the social pressures of their peers. Thus, reddit may be the conduit for investors to convince each other to behave like a herd. A caveat worth mentioning is the initial source of the herd behavior does not have to be itself enormous to start a large rush. The /r/wallstreetbets community only reached one million subscribers in 2020 (SubredditStats 2021), and the daily volume of trades for a popular stock like Tesla can easily exceed ten million. However, even small initial movements in the price of a stock can incite momentum traders and large funds to take note. During the GameStop bubble, large established hedge funds were

actually following in the footsteps of the retail traders (MacMillan, 2021). Political figures, like Congressperson Alexandria Ocasio-Cortez, and business leaders, like Tesla CEO Elon Musk, used their platforms to publicize the GameStop bubble, further drawing attention and investment to the stock.

Prior research that looks at connections between social media and markets uses machine learning to process information embedded in user-generated chatter, and look at asset prices to determine their relationship. I will be using unsupervised clustering to classify Reddit posts, and using those to predict the daily closing prices of five stocks. This paper is novel in that it uses unsupervised k-means clustering for classification, instead of sentiment analysis like past work, and this is the first study to examine Reddit comments. In addition, I will be looking at frequency of individual terms, and not just broad measures of sentiment.

## 3. Literature Review

The core assumption of homogenous, perfect information among all investors does not hold up to scrutiny. New events that may affect the price of an equity are constantly happening. Nearly every asset is affected by systematic events like the actions of the Federal Reserve or major financial crises. Some events are industry specific. For example, defense stocks in particular rose after the killing of Iranian general Qassem Suleimani and heightened US-Iranian tensions in January 2020. Events that influence stock price may be extremely firm-specific, like when a CEO dies (Johnson, Magee, Nagarajan & Newman, 1985), or a product is granted regulatory approval (Sturm, Dowling & Roder 2007). The addition of a stock to an index like the S&P 500 can cause its price to increase (Lynch & Mendenhall 1997), as ETFs and mutual funds that track that index switch their capital to the newly added stock.

However, there are costs to information. Researching and modeling everything that can influence the price of a security takes time, effort and skill, as indicated by the six-figure salaries that analysts can command. At the same time, there are diminishing returns to informed trading, as once every market participant is perfectly informed, there is no reason to trade; prices have already incorporated all information. What this means is that, in equilibrium, only a certain fraction of investors will be well-informed (Stiglitz & Grossman 1976). Sophisticated and unsophisticated investors will coexist.

Luckily, there are other sources of information. The media can sensationalize, but journalists are incentivized to be accurate. A financial news source like the Wall Street Journal or Financial Times would lose credibility and revenue for consistently spreading misinformation (Schuster 2003). Ergo, third-party sources can be reliable. Investors themselves may have incentives to participate in communities where they reveal their private information or trading strategies to each other. One theory is the diversification of risky strategies: if investors share their ideas with one another, they can diversify their portfolios at the cost of slightly lower returns (Gray 2010). Within these communities, those who share information are rewarded socially, with high reputation and esteem among their peers (Park, Gu, Leung & Konana 2014). Given this, internet communities may be repositories of useful market information. An example of this is shown in Mizrach & Weertz 2009, which documents the users of activetrader, an online forum for individual traders. Users of this site would publicly share and discuss their positions. Per the analysis by the authors of the paper, many of these traders may have been very successful.

A tweet, comment or blog post is by nature a very unstructured, qualitative piece of data. The field of natural language processing is devoted to turning text into numbers. A simple approach is to turn a document into a "bag of words" (Zhang, Jin & Zhou 2010). First, it must be tokenized into single words, or n-grams (n words in a series, like "I love you"). Then each tweet can be represented as a frequency vector V, where v_i represents the relative frequency of word or n-gram i, in the given tweet. The most basic approach is to preselect some words or n-grams of interest, and only look at those. Bollen, Mao & Zeng 2011 used 964 different terms, each with a weight associated with six different emotional states: calm, sure, alert, vital, kind and happy, in addition to 7630 terms that were marked as positive or negative. This weight assumes that terms have the same weight and meaning in different contexts, which is not the case. Gholampour & van Wincoop 2017 performed a similar task, but they levied 10 financial professionals to rate certain terms as "positive" or "negative", creating a finance specific lexicon of words with polarity scores.

Instead of trying to find the sentiment of individual terms, some have cut out the middle man, classifying the entire document without worrying about the polarity of the terms or groups of terms themselves. Classification is done using supervised machine learning, where the parameters of the model are trained on a subset of pre-labeled documents. This requires researchers to not only manually label hundreds, or even thousands of documents, and then decide what features to use as inputs to the model, as well as what model to use for classification. Using only a vector of term frequencies to represent each document, similar to Hamilton, Clark, Leskovec & Jurafsky 2016, one can train a model like a Support Vector Machine (Ranco et al 2015), Random Forest (Dridi, Atzeni & Recupero 2017), or Artificial Neural Network (Ghiassi, Skinner & Zimbra 2013), which can also be used to classify documents into more than two categories. Dridi, Atzeni & Recupero 2017 also added semantic features, such as the actual meanings of the words, and the presence of synsets, sets of words with similar meetings, all predefined and not domain-specific as inputs to their model. Alternatively, one could use a slightly different label-propagation algorithm, where documents are nodes, and edges between documents are the cosine similarity between their row vectors (Speriosu, Upadhyay, Sudan & Baldridge 2011), which experimentally outperformed a traditional Maximum Entropy classifier and a lexicon-based classification dictionary.

Studies have looked at the correlation between text data and indices such as the Dow Jones (Bollen, Mao & Zeng 2011), individual stocks, like those in the Dow Jones (Ranco et al 2015) or Argentina's leading index (Galvez & Gravano, 2017), or exchange rates (Gholampour, van Wincoop 2017). In addition to looking at prices, researchers have also looked at the behavior of short sellers in response to news (Engelberg & Reed Ringgenberg 2010), or the relative behavior of investors who are considered technical or non-technical investors (Lopez-Cabarcos, Piniero-Chousa & Perez-Pico 2017). Daily lexical data, like sentiment can be used as inputs for classification, classifying each day as a day where the price increases or not, using a model like Random Forest or a Fuzzy Neural Network (Galvez & Gravano, 2017). Causality can be tested for between time-series of daily sentiment and changes in financial variables (Gholampour & van Wincoop 2017). Ranco et al used an event study, where "events" were single days with abnormal jumps in tweets that mentioned specific stocks.

This paper expands on previous NLP by combining the Latent Semantic Indexing (LSI) for dimensionality reduction and K-Nearest Neighbors (KNN) , also known as K-Means Clustering, for unsupervised classification, in a manner similar to Chen, Xiao, Sheng, & Teredesai 2017. LSI is a technique that compresses the information of a high-dimensionality vector into a lower-dimensionality vector. KNN is a technique that groups n-dimensional vectors, each of which represents an observation to be classified, into clusters of vectors that are most similar to each other (Cheng 1984). I will study Reddit comments that contain mentions of large publicly traded stocks, and use the frequency of classes, as determined by the KNN algorithm. Like past studies, I will use daily returns as the dependent variable.

## 4. Data

### Financial Data

I am looking at five stocks, Microsoft (MSFT), Walmart (WMT), Nike (NKE), Tesla (TSLA) and Pfizer (PFE). All were chosen because they are large, well-known corporations that there would be significant social media discussion of. In addition, none of their names are homonyms. When parsing comment data, I would have to use contextual clues to determine whether a reference to Caterpillar refers to the insect or the stock. For each stock, I also found an industry-tracking exchange traded fund (ETF) that should perform similarly to it, as shown in the following table. Tesla was hard to categorize, as it manufactures semi-autonomous electric sports cars. It is not a traditional tech company like Google, a traditional automaker like Ford or typical luxury goods merchant like Tiffany's. While I initially wanted to use an electric vehicle ETF, Bloomberg had missing data for some of the days of the period in question, so I settled on the Lithium industry, as Tesla's vehicles all use lithium ion batteries. Over the period between January 1st 2018 to December 31st, 2019, dropping days where no observations were available, the correlation between the daily returns of Tesla and the First Trust NASDAQ Global Auto Index ETF (CARZ), respectively was 0.38. For the same time period, the correlation between the daily returns of iShares U.S. Technology ETF (IYW) and Tesla was 0.4. The correlation between daily Tesla returns and daily Lithium & Battery Tech ETF (LIT) returns over the same period was 0.41.

**Table 1:**

| **Industry** | **Stock Symbol** | **ETF Symbol** |
|---|---|---|
| Retail | WMT | XRT |
| Technology | MSFT | IYW |
| Consumer Non-Durables | NKE | VDC |
| Pharmaceuticals | PFE | XPH |
| Lithium Industry | TSLA | LIT |

**Social Media Data**

For this study, I parsed Reddit for comments that mentioned any of the selected stocks' common names or stock symbol. Reddit is a discussion board website, divided into subreddits, which are discussion boards for particular topics. I.e, the news subreddit (reddit.com/r/news) is for discussing news, but there are stranger, niche topics like reddit.com/r/birdswitharms. The advantage of this subreddit structure is that one can be reasonably sure that in the context of a subreddit with a financial topic, users will be discussing the markets or equities. For this particular study, I looked at only one subreddit, known as “/r/wallstreetbets”. This subreddit is known for its crude and vulgar culture, with users frequently using profanity and slurs. However, that does not necessarily mean that “/r/wallstreetbets” users are uninformed or even unintelligent. Comments on this subreddit can be very clear forecasts, like “Bless you TSLA! Made almost 8 grand this morning, still not selling my puts until friday; I think it has farther to fall.” (Comment id ed2vs1p), to the almost nonsensical if you do not know the modern internet language: “big oof TSLA” (Comment id ed2ve4t). An advantage “/r/wallstreetbets” is that members often post pictures of their portfolios and trades, which as far as I know is not common in any other virtual community. This leads me to theorize that the members of this community are not just “armchair investors” but active traders, who are actually trading, and possibly learning from the mistakes of themselves and their peers.

In order to actually retrieve the comments, I used the Pushshift API, a massive publicly available dataset of reddit posts, comments and user statistics from 2005 to (as of writing this) the end of 2019 (Baumgartner, Zannettou, Keegan, Squire & Blackburn 2020). Using the API, I requested every comment posted to “/r/wallstreetbets” from June 1st, 2018, to February 21st, 2019. As the API limited me to a maximum of 5000 comments per request, I had to break up the entire 367-day time period into consecutive 50 minute intervals. If there were more than 5000 comments within one 50-minute interval, I would send five more requests, corresponding to every ten-minute epoch within the 50-minute interval. I then saved the unique ids, timestamp, and text content of every comment returned that contained mentions of a relevant stock being studied, and otherwise only saved the unique ids, timestamp and. Special characters (like emojis) and accented characters were removed, and all text was converted to lowercase.

Once the data is collected, for each stock, I can construct a frequency statistic, like so:

$$f_{all,t} = \frac{\text{quantity of comments that mention stock i on day t}}{\text{quantity of comments on day t}}$$

Thus creating a daily time series of frequencies for each asset. Then, I can classify comments, and create a class-stock frequency statistic like so:

$$f_{c,t} = \frac{\text{quantity of comments classified as class c that mention stock i on day t}}{\text{quantity of comments on day t}}$$

This will create for each stock, for each class, a daily time series. The quantity of classes will depend on the comment classification algorithm.

For both of these, I will square the daily quantities to see if there are increasing or decreasing effects, creating frequency squared time series:

$$f^2_{all2,t} = \frac{(\text{quantity of comments that mention stock i on day t})^2}{\text{quantity of comments on day t}}$$

$$f_{c2,t} = \frac{(\text{quantity of comments classified as class c that mention stock i on day t})^2}{\text{quantity of comments on day t}}$$

## 5. Methodology

### Classification

The process of classification is broken up into three parts: vectorization, dimensionality reduction and clustering.

Vectorization is the process of turning a corpus of documents (in this case, Reddit comments) into a bag-of-words and n-grams. Usually, stopwords, like “the”, “no” and “but”, are excluded from this vector. I used the built-in set of stopwords provided by the Scikit-Learn library, which compromises 318 english words. The remaining terms may be weighted via the Term Frequency- Inverse Document Frequency (TF-IDF) weighting, which gives higher weights to terms that appear very frequently in a small number of documents (Wikipedia 2020). For a term t, in document d, the raw Term Frequency (TF) would be:

$$tf_{t,d} = \frac{countoftermtindocumentd}{countofalltermsindocumentd}$$

The Inverse Document Frequency (IDF) would be, for term t, assuming there are n documents in the corpus:

$$idf(t) = log[n/df(t)] + 1$$

While the TF-IDF would be, for term t in document d:

$$tf - idf(t,d) = tf(t,d) * idf(t)$$

Dimensionality reduction is the process of turning high-dimensional sparse vectors that represent data, like documents, into compact low-dimensionality representations that preserve most of the information in the high-level representations. This is a form of feature extraction.

The advantage of dimensionality reduction is that it improves computational cost of the K-Nearest Neighbors (KNN) algorithm I will be using for clustering, and sometimes improves accuracy (Gayathri & Marimuthu 2012). this research project, I will be using Latent Semantic Indexing (LSI), Given m terms or features, and and n documents, we turn each document into a j-dimensional vector, where j is a chosen hyperparameter, j <<< m, that preserves relationships and cooccurrences between similar words (Berry, Dumais & G.W. O'Brien 1995), thus minimizing information loss. We construct the m x n matrix A (where columns represent documents and ,rows represent terms) which can be decomposed like so:

$$A = U\Sigma V^T$$

The diagonals of $\Sigma$ are the singular values, and each column of $V^T$, down to the jth row, is a j dimension approximation of the corresponding column in A.

Clustering is just grouping similar data points.The advantage of clustering is that we do not need a training set of pre-categorized comments to cluster them, so we can classify documents without supervision. KNN is a technique that groups n-dimensional vectors, each of which represents an observation to be classified, by randomly determining “centroids” that are vectors that are close (under some distance metric like cosine similarity or euclidean distance). It eventually assigns each document to one of k clusters (Cheng 1984).

Both k and j are chosen to find the k and j combination that maximizes the average silhouette of each cluster (Kodinariya & Makwana 2013). The silhouette metric of each cluster is the average similarity of each member of the cluster to other members of the cluster relative to the average similarity of each member of the cluster to data points not in the cluster (Rousseuw 1987). It ranges from 0, which indicates that the clusters are pretty much all the same, to 1, which indicates that clusters are very distinct.

| Stock | Optimal J | Optimal K | Average Silhouette |
|---|---|---|---|
| WMT | 4 | 4 | 0.58 |
| MSFT | 4 | 3 | 0.6 |
| NKE | 4 | 3 | 0.56 |
| PFE | 4 | 3 | 0.59 |
| TSLA | 4 | 5 | 0.63 |

**Visualization**

Each group of comments will have different frequencies of different terms. To demonstrate, in the following wordclouds, a larger word implies a greater frequency of the word in the comment group.

**I. Walmart**

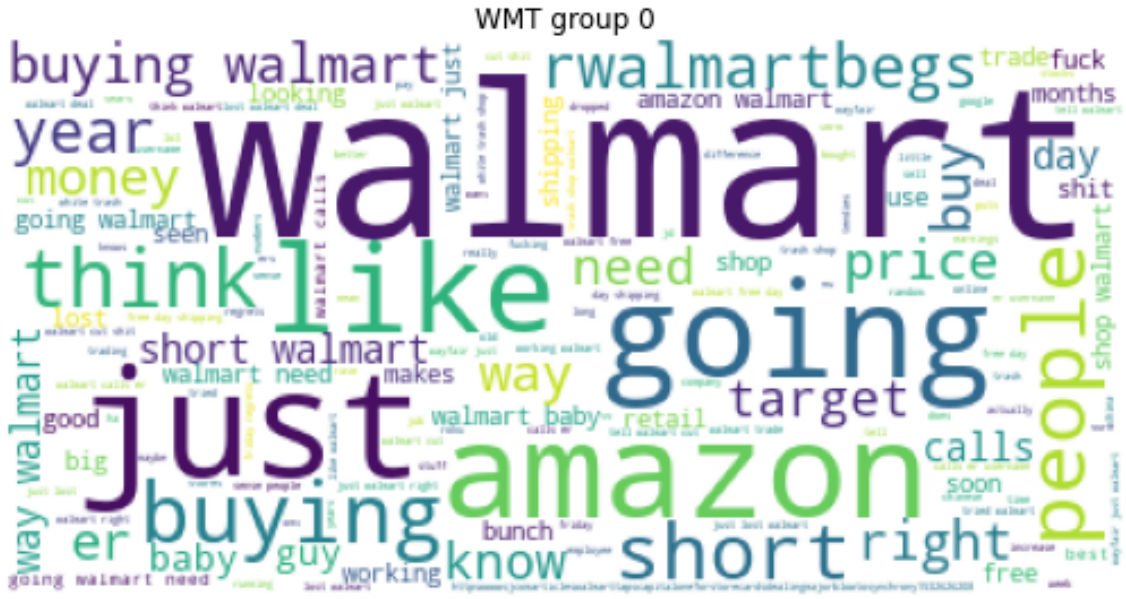

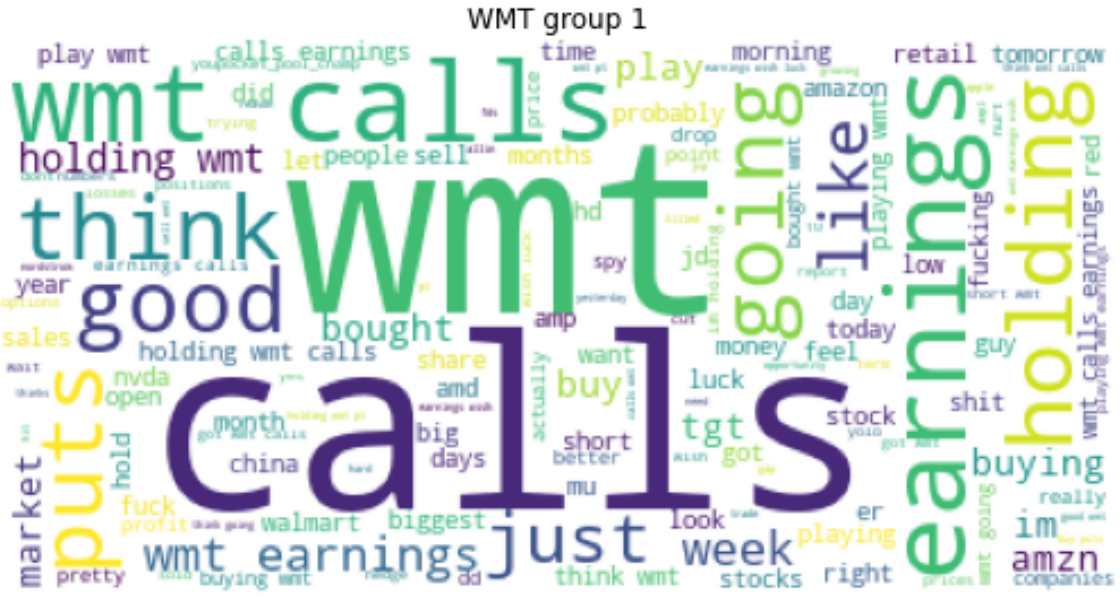

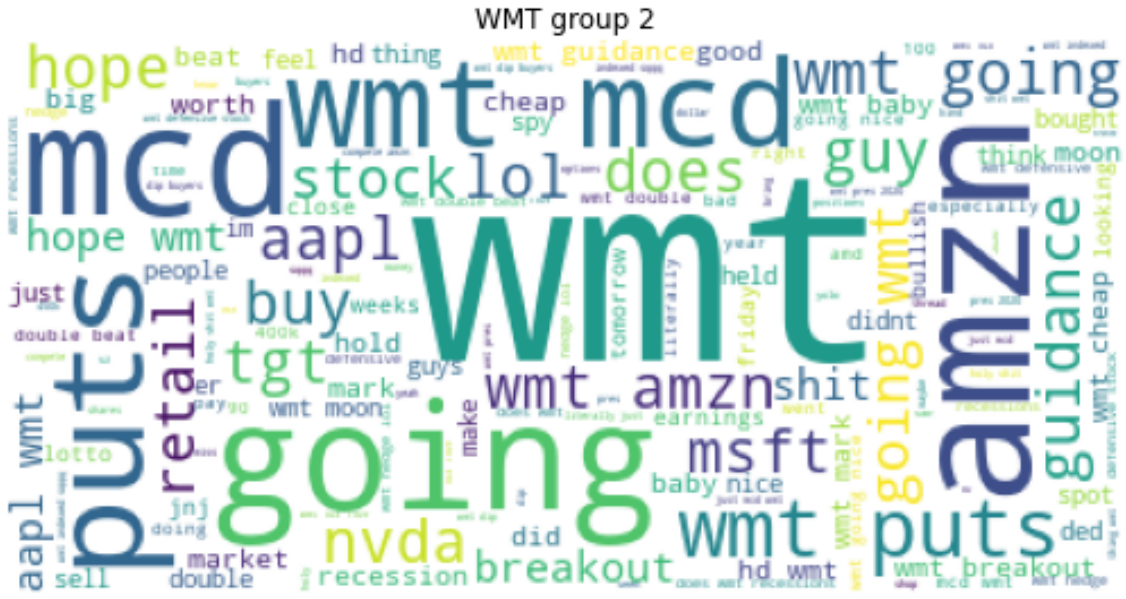

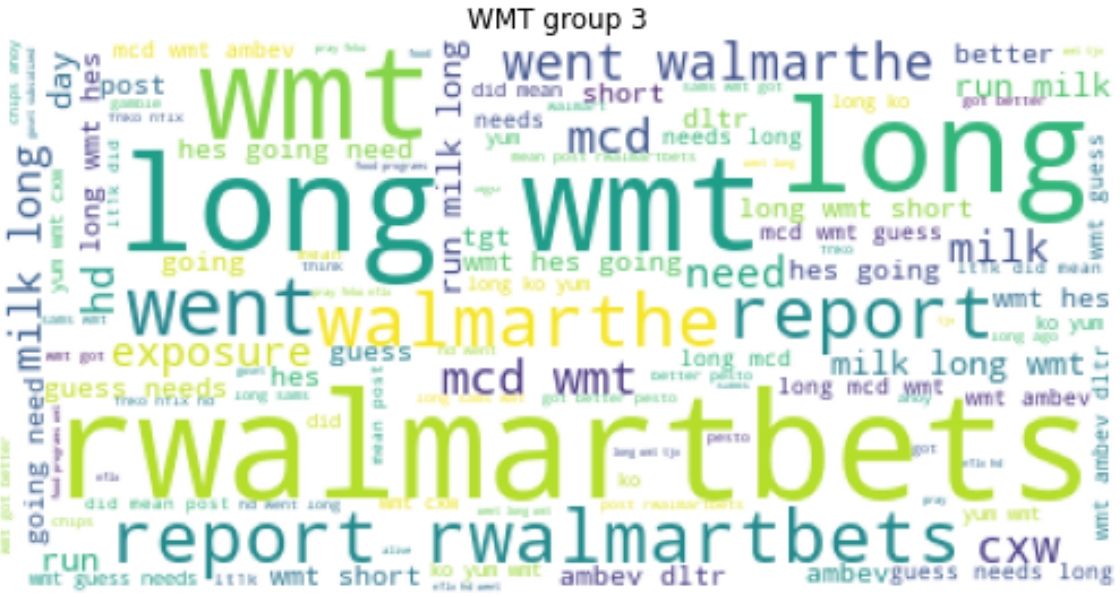

**II. Microsoft**

MSFT group 0

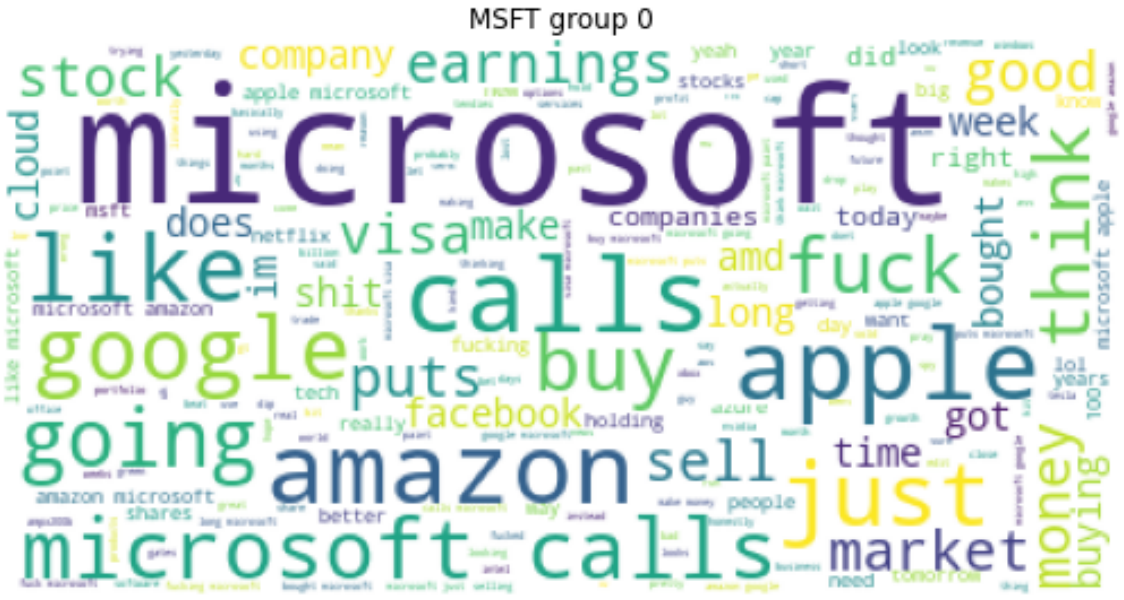

MSFT group 1

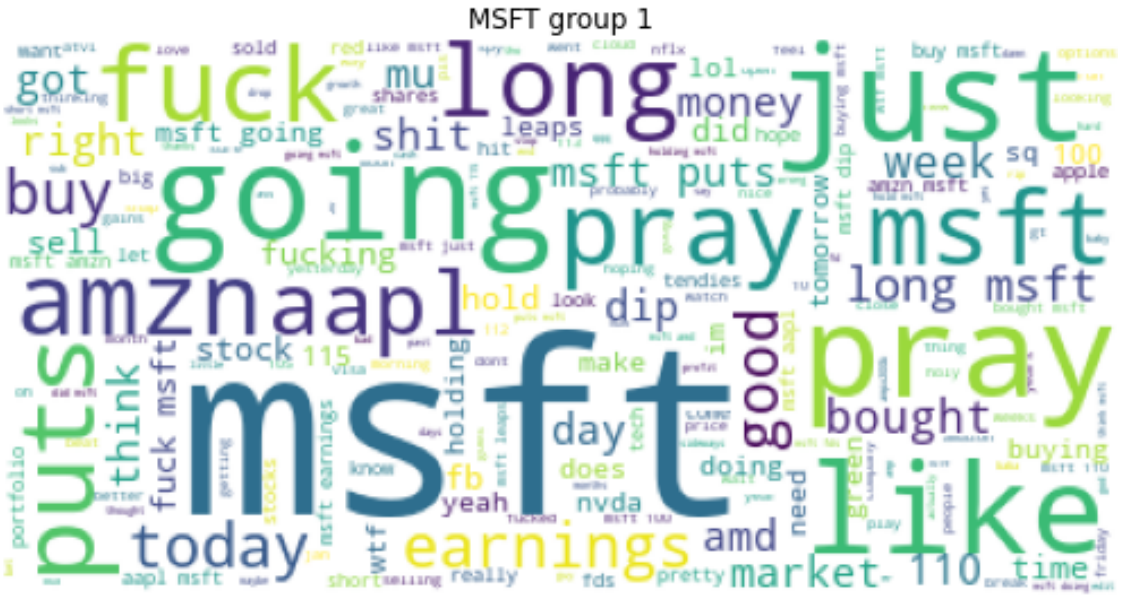

MSFT group 2

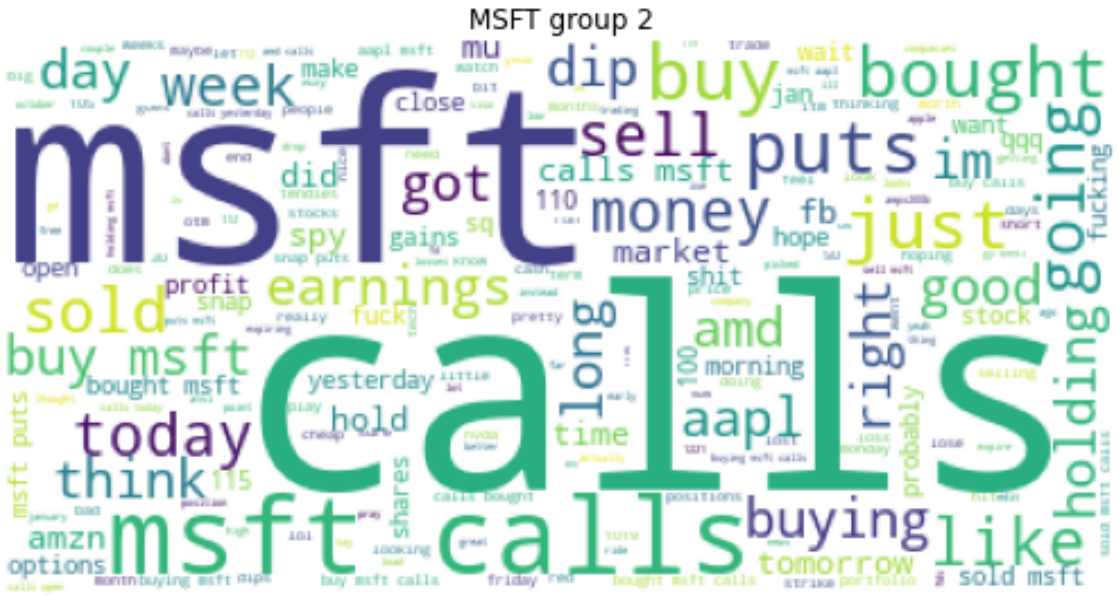

# III. Nike

NKE group 0

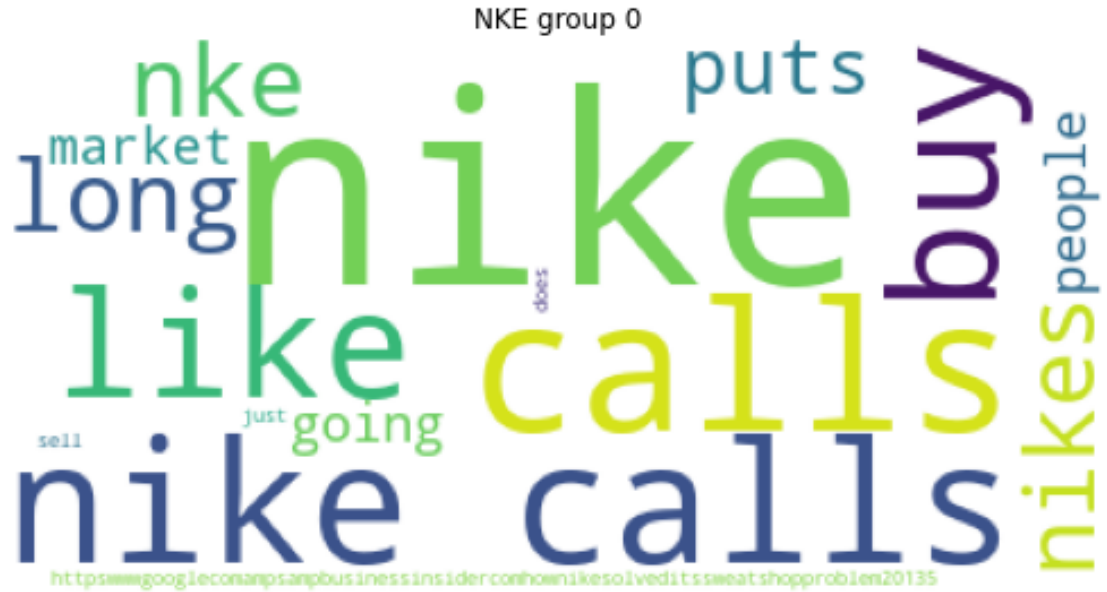

NKE group 1

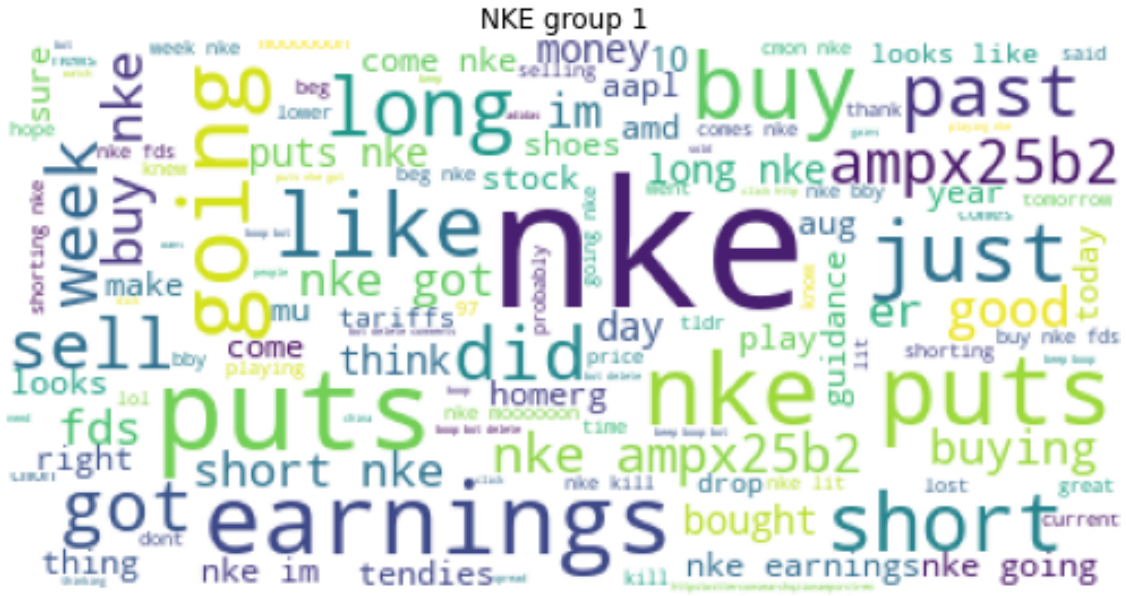

NKE group 2

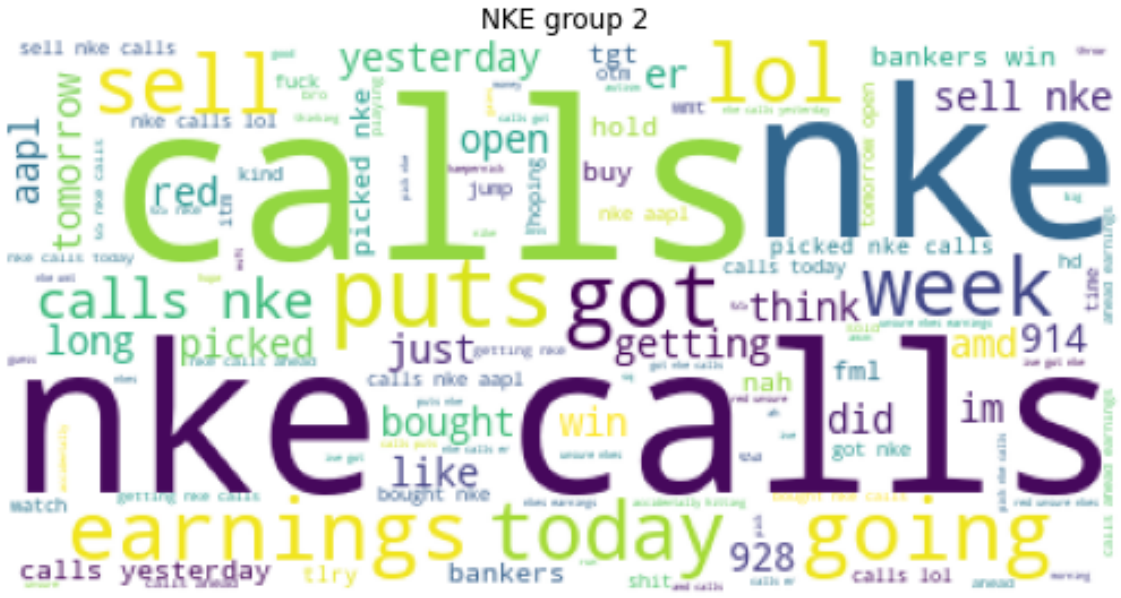

# IV. Pfizer

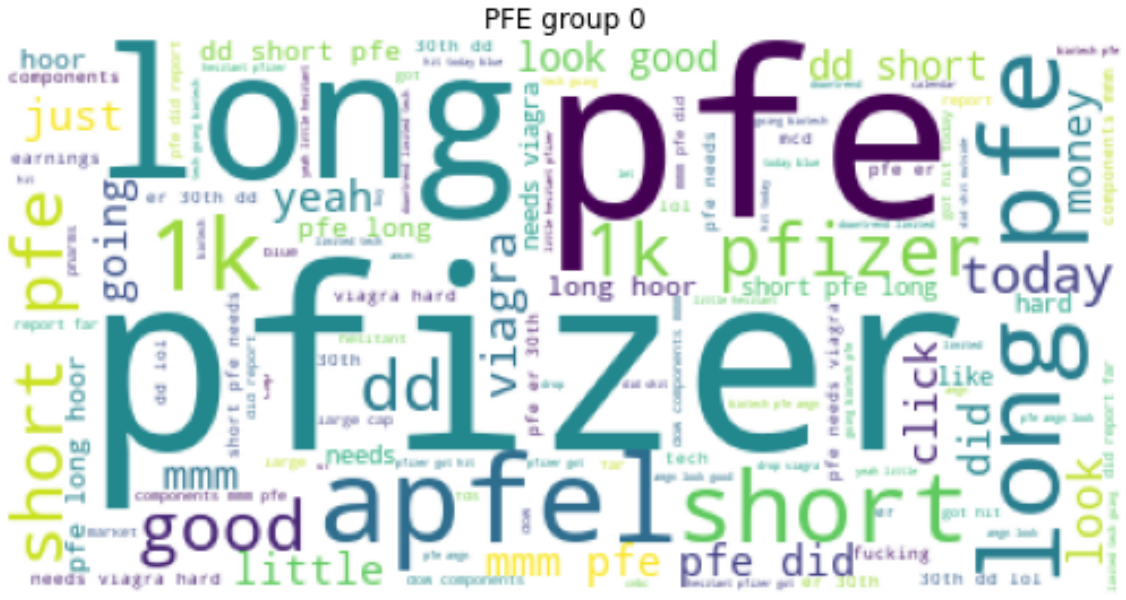

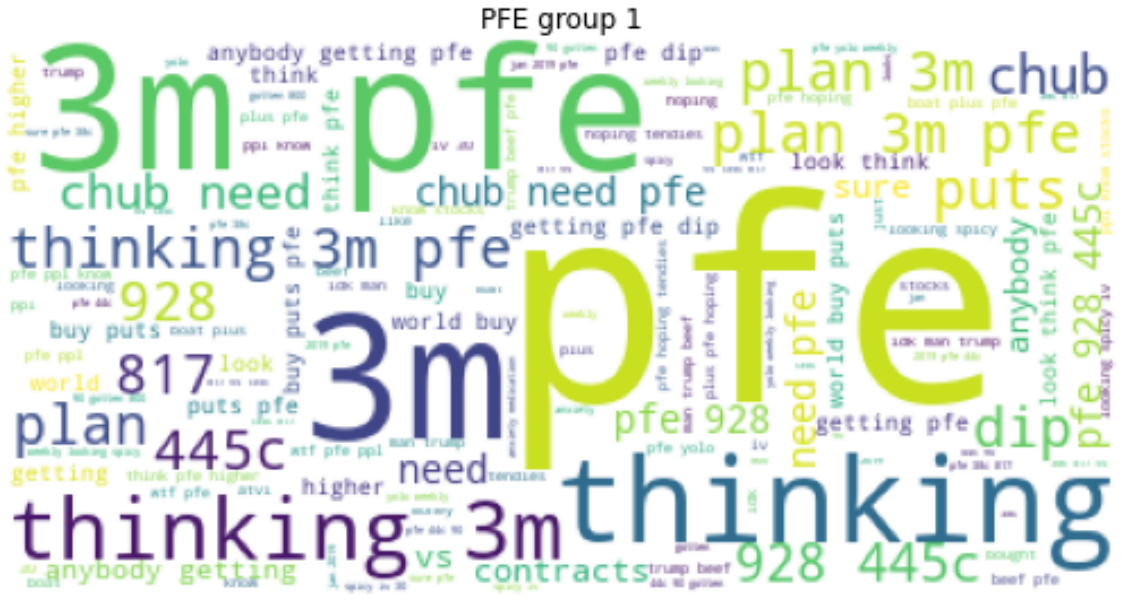

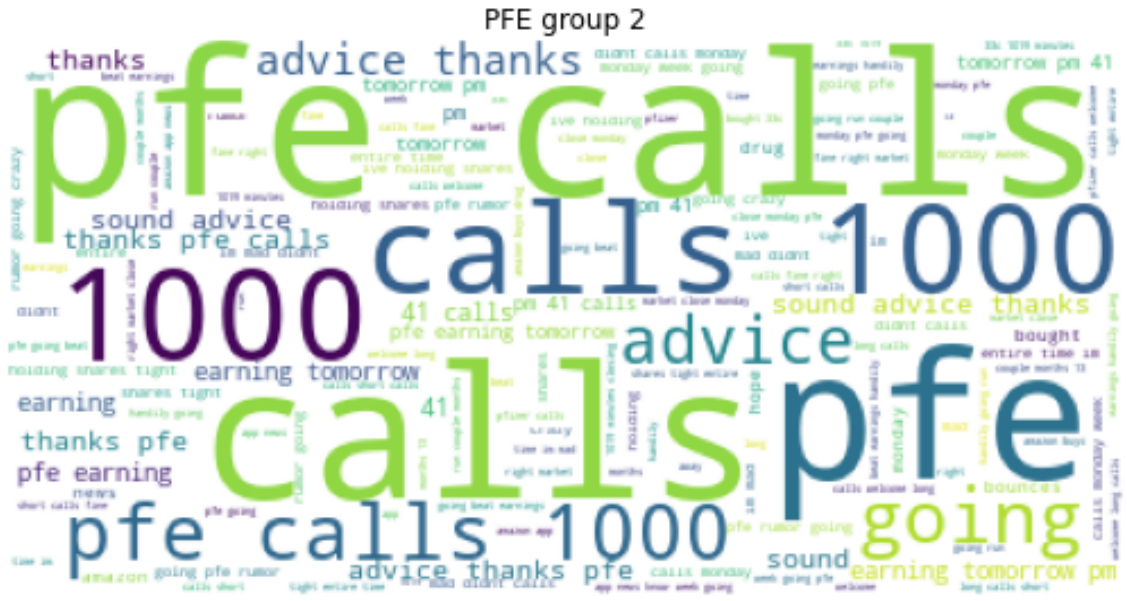

## V. Tesla

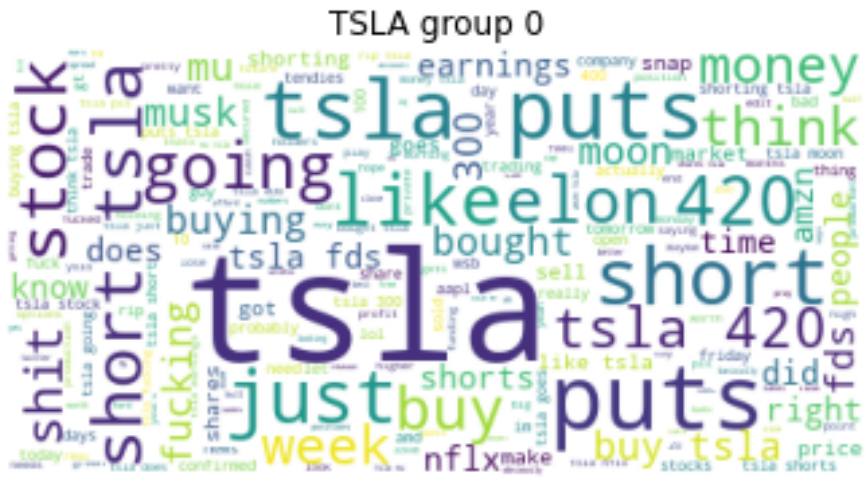

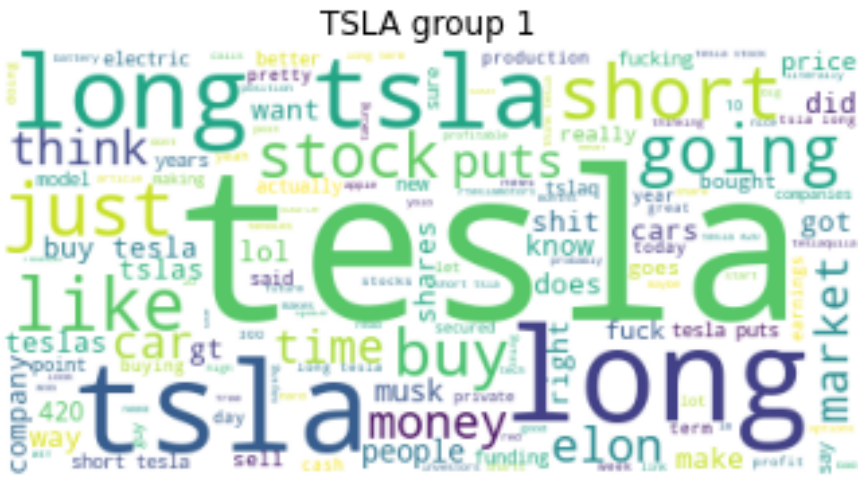

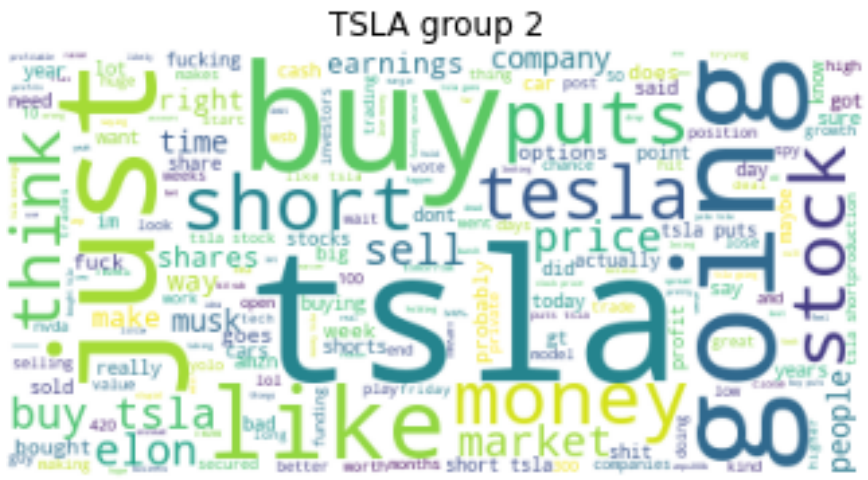

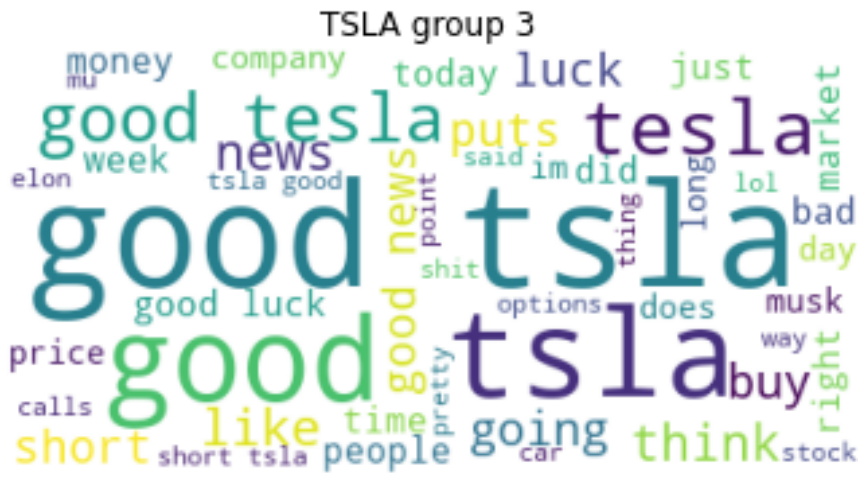

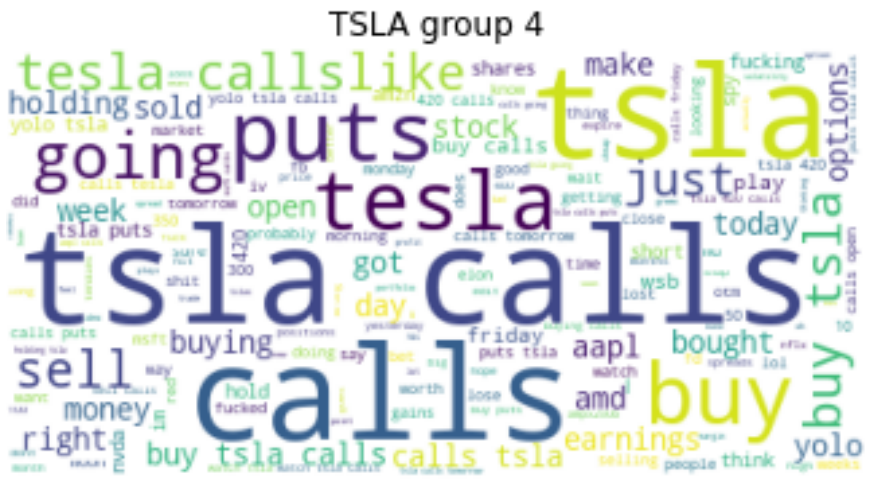

Notably, many of the groups share some of the same words, but with different emphasis.

**Buzzwords**

I also looked at the daily frequencies of "buzzwords" for each stock. This is similar to approaches like Fung, Yu & Lu (2005), where the features used to predict a news story's impact on stocks are the presence of words in the story. The buzzwords were selected by finding the 100 most frequent n-grams for each corpus of comments belonging to a specific stock. Then of those 100, I selected the 10 with the highest magnitudes of correlation between the daily returns and the frequencies over the time period. I divided the daily count of each word by the daily volume of comments on /r/wallstreetbets. On a particular day:

$$w_c = \frac{\text{mentions of c}}{\text{daily comment volume}}$$

The buzzwords for each stock were as following:

| | |
|---|---|
| WMT | 'earnings', 'lol', 'going', 'wmt earnings', 'tgt', 'nvda', 'buy', 'puts', 'wmt going', 'play' |
| MSFT | 'fuck', 'microsoft', 'buying', 'right', 'fucking', 'stock', 'think', 'tomorrow', 'amzn', 'buy msft' |
| NKE | 'watch', 'earnings', 'make', 'nke calls', 'calls', 'did', 'people', 'short', 'shoes', 'long nke' |
| PFE | 'long pfe', 'long', 'need', 'short pfe', 'just', 'good', 'short', 'buy', 'dd', 'pfe calls' |
| TSLA | 'good', 'musk', 'lol', 'elon', 'car', 'company', 'does', 'time', 'mu', 'earnings' |

**Fama-French Three-Factor Model**

The Capital Asset Pricing Model (CAPM) is a model for modeling the expected return of a risky asset as a function of a risk free interest rate and expected return of a portfolio of the universe of risky assets, when the assets have been weighted to minimize variance (Fama, French 2004). For asset i, in a universe of N risky equities, the expected return:

$$E[R_i - R_f] = \alpha_i + \beta_{i,1} E[R_M - R_f], i = 0, 1, 2, , , , N$$

$R_f$ is the risk-free interest rate, $E[R_M - R_f]$ is the return on the entire minimum variance portfolio that represents the entire universe of risky assets minus the risk free rate (Fama & French 2004). In practice, I will use the daily returns of the overall market or an industry specific ETF for this.

The Fama-French three factor model is an extension to the CAPM pricing model, which incorporates two more factors. The High-Minus-Low (HML) factor is the returns of a high book-to-equity portfolio minus that of a low book-to-equity portfolio. Essentially, it reflects the premium to value stocks over growth stocks. The Small-Minus-Big (SMB) factor is the returns of a small-capitalization stock portfolio minus that of a large-capitalization stock portfolio, and it represents the premium to smaller firms.

$$E[R_i - R_f] = \alpha_i + \beta_{i,1}(R_M - R_f) + \beta_{i,2}HML + \beta_{i,3}SMB$$

**Extending the Fama-French with Reddit Factors**

For each stock, for each benchmark (The broader market or a specific ETF), I will test five models, using the frequencies and frequencies squared:

$$E[R - R_f] = \alpha_i + \beta_1(R_M - R_f) + \beta_2 HML + \beta_{i,3}SMB + \beta_{all} f_{all,t}$$

$$E[R - R_f] = \alpha_i + \beta_1(R_M - R_f) + \beta_2 HML + \beta_3 SMB + \sum_{c=0}^{k} \beta_{f,c} f_{c,t}$$

$$E[R - R_f] = \alpha_i + \beta_1(R_M - R_f) + \beta_2 HML + \beta_3 SMB + \beta_{fall} f_{all,t} + \beta_{all2} f^2_{all2,t}$$

$$E[R - R_f] = \alpha_i + \beta_1(R_M - R_f) + \beta_2 HML + \beta_3 SMB + \sum_{c=0}^{k} \beta_{f,c} f_{c,t} + \sum_{c=0}^{k} \beta_{f,c2} f_{c2,t}$$

$$E[R - R_f] = \alpha_i + \beta_1(R_M - R_f) + \beta_2 HML + \beta_3 SMB + \sum_{c=0}^{10} \beta_{w,c} w_c$$

In addition, for each of these regressions, I lagged each reddit variable by one day, doubling the number of models fit. The regression results are shown in the output section at the end of this paper. To test these models, we perform a hypothesis test, where the null hypothesis is that all the coefficients of the added variables are 0 using the F-statistic.

## 6. Regression Results

The following tables display the regression results. Each table corresponds to a model fit for the daily returns of one particular stock, and each row represents an independent variable. In Table 6.a.i, for example, the value of -0.66 in the row SMB, and the column PFE, means the beta coefficient on the SMB in a regression with the daily returns of Pfizer (PFE) is -0.66. The final row is the value of the F-statistic derived from comparing the complete model to a nested model that only uses the three French-Fama factors (Rm, HML and SMB). A significant F means that we reject the null hypothesis that the coefficients on the extra factors are zero. The significance of coefficients/F-statistics is indicated by the asterisks following them. One asterisk means significance at the 0.1 level, two means significance at the 0.05 level, and three asterisks means significance at the 0.01 level.

**a. French-Fama with sums of all comments:**

$$E[R - R_f] = \alpha_i + \beta_1(R_M - R_f) + \beta_2 HML + \beta_{i,3} SMB + \beta_{all} f_{all,t}$$

**i. Using the Market as a benchmark:**

| | WMT | MSFT | NKE | PFE | TSLA |
|---|---|---|---|---|---|
| alpha | -0.00 | 0.13 | 0.00 | 0.11 | 0.02 |
| Rm | 0.52*** | 1.03*** | 0.65*** | 0.67*** | 0.75*** |
| HML | 0.16 | -0.12 | -0.93*** | -0.04 | -1.11** |
| SMB | -0.56*** | -0.26** | 0.23 | -0.66*** | -0.25 |
| $f_{all}$ | 3602.75 | -108.66 | -3393.04 | -162.46 | 1052.98 |
| F | 0.46 | 0.38 | 0.67 | 0.00 | 0.12 |

**Table 6.a.i**

**ii. Using the ETF as a benchmark:**

| | WMT | MSFT | NKE | PFE | TSLA |
|---|---|---|---|---|---|
| alpha | 0.11 | 0.05 | 0.10 | 0.15* | 0.09 |
| Rm | 0.58*** | 1.24*** | 1.03*** | 0.76*** | 0.86** |
| HML | 0.38** | -0.82*** | -0.25 | 0.14 | -0.99 |
| SMB | -0.12 | -0.44*** | 0.15 | -0.40** | -0.23 |
| $f_{all}$ | -3414.59 | -63.57 | -2683.21 | 7140.88 | 305.02 |
| F | 0.38 | 0.10 | 0.53 | 0.13 | 0.01 |

**Table 6.a.ii**

**b. French-Fama with categorized comments:**

$$E[R - R_f] = \alpha_i + \beta_1(R_M - R_f) + \beta_2 HML + \beta_3 SMB + \sum_{c=0}^{k} \beta_{f,c} f_{c,t}$$

**i. Using the Market as a benchmark:**

| | WMT | MSFT | NKE | PFE | TSLA |
|---|---|---|---|---|---|
| alpha | -0.03 | 0.13 | -0.03 | 0.11 | -0.22 |
| Rm | 0.54*** | 1.03*** | 0.64*** | 0.67*** | 0.81*** |
| HML | 0.17 | -0.12 | -0.96*** | -0.05 | -1.41** |
| SMB | -0.56*** | -0.26* | 0.19 | -0.66*** | -0.18 |
| $f_0$ | 23231.91* | -43.50 | -65088.00** | 10221.71 | 75563.55*** |
| $f_1$ | 3192.10 | -401.70 | 11535.70 | 7712.97 | 14701.52 |
| $f_2$ | -12405.05 | -57.48 | 47349.88 | -62082.66 | -33970.56** |
| $f_3$ | -31863.05 | | | | 17184.84 |
| $f_4$ | | | | | -42662.35 |
| F | 0.96 | 0.14 | 1.87 | 0.29 | 2.78** |

**Table 6.b.i**

**ii. Using the ETF as a benchmark**

| | WMT | MSFT | NKE | PFE | TSLA |
|---|---|---|---|---|---|
| alpha | 0.08 | 0.06 | 0.07 | 0.15* | -0.14 |
| Rm | 0.58*** | 1.21*** | 1.00*** | 0.75*** | 0.86** |
| HML | 0.38** | -0.82*** | -0.29 | 0.13 | -1.28** |
| SMB | -0.11 | -0.44*** | 0.12 | -0.40** | -0.17 |
| $f_0$ | 11645.57 | -134.22 | -34931.29 | 14254.34 | 76102.12*** |
| $f_1$ | -69.80 | -1874.91 | 3056.15 | 18385.92 | 8379.67 |
| $f_2$ | -23158.65 | 431.80 | 33915.84 | -47529.92 | -32371.81** |
| $f_3$ | -14218.72 | | | | 17083.68 |
| $f_4$ | | | | | -34599.03 |
| F | 0.54 | 0.60 | 0.90 | 0.27 | 2.45** |

**Table 6.b.ii**

**c. French-Fama with sums of all comments and squared sums:**

$$E[R-R_f] = \alpha_i+\beta_1(R_M-R_f)+\beta_2 HML+\beta_3 SMB+\beta_{fall} f_{all,t}+\beta_{all2} f_{all2,t}^2$$

**i. Using the Market as a benchmark:**

| | WMT | MSFT | NKE | PFE | TSLA |
|---|---|---|---|---|---|
| alpha | -0.07 | -0.11 | -0.12 | 0.12 | -0.07 |
| Rm | 0.52*** | 1.02*** | 0.71*** | 0.67*** | 0.75*** |
| HML | 0.16 | -0.13 | -0.94*** | -0.03 | -1.13** |
| SMB | -0.56*** | -0.22* | 0.22 | -0.64*** | -0.24 |
| $f_{all}$ | 21692.16* | 1170.17* | 16891.79 | -48597.83 | 2351.02 |
| $f_{all}^2$ | -6795.85* | -19.02* | -1701.58** | 34496.59 | -85.63 |
| F | 1.75 | 1.94 | 2.35* | 0.26 | 0.07 |

**Table 6.c.i**

**ii. Using the ETF as a benchmark:**

| | WMT | MSFT | NKE | PFE | TSLA |
|---|---|---|---|---|---|
| alpha | 0.05 | -0.12 | 0.00 | 0.17* | 0.03 |
| Rm | 0.57*** | 1.22*** | 1.03*** | 0.75*** | 0.86** |
| HML | 0.38** | -0.82*** | -0.26 | 0.15 | -1.00 |
| SMB | -0.12 | -0.41*** | 0.13 | -0.38** | -0.23 |
| $f_{all}$ | 14968.06 | 843.30 | 13546.97 | -47887.73 | 1086.37 |
| $f_{all}^2$ | -6874.54* | -13.51 | -1368.86* | 39198.24 | -51.50 |
| F | 1.60 | 0.72 | 1.95 | 0.39 | 0.01 |

**Table 6.c.ii**

**d. French-Fama with categorized comments and comments squared:**

$$E[R-R_f] = \alpha_i + \beta_1(R_M - R_f) + \beta_2 HML + \beta_3 SMB + \sum_{c=0}^{k} \beta_{f,c} f_{c,t} + \sum_{c=0}^{k} \beta_{f,c2} f_{c2,t}$$

**i. Using the Market as a benchmark:**

| | WMT | MSFT | NKE | PFE | TSLA |
|---|---|---|---|---|---|
| alpha | -0.10 | -0.14 | -0.20 | 0.12 | -0.01 |
| Rm | 0.54*** | 1.01*** | 0.68*** | 0.67*** | 0.88*** |
| HML | 0.19 | -0.14 | -0.95*** | -0.05 | -1.29** |
| SMB | -0.55*** | -0.22 | 0.20 | -0.64*** | 0.09 |
| $f_0$ | 39595.15 | 4798.75 | -43130.85 | -44916.85 | 97980.48 |
| $f_1$ | 46758.58** | 1800.76 | 30862.22 | 2085.12 | 26062.80 |
| $f_2$ | -24187.29 | -92.82 | 129334.87** | -29928.12 | -24363.89 |
| $f_3$ | -20011.06 | | | | -63531.27** |
| $f_4$ | | | | | 144251.04 |
| $f_0^2$ | -9364.02 | -100.89 | 2091.93 | 33406.33 | -3334.41 |
| $f_1^2$ | -12366.09** | -22.36 | -1768.38 | 2085.12 | 722.88 |
| $f_2^2$ | 4229.77 | -9.43 | -31618.09** | -29928.12 | -1062.61 |
| $f_3^2$ | -20011.06 | | | | 4515.21*** |
| $f_4^2$ | | | | | -48303.67* |
| F | 1.32 | 0.97 | 2.04* | 0.30 | 2.98*** |

**Table 6.d.i**

**ii. Using the ETF as a benchmark**

| | WMT | MSFT | NKE | PFE | TSLA |
|---|---|---|---|---|---|
| alpha | 0.02 | -0.07 | -0.09 | 0.16* | 0.24 |
| Rm | 0.59*** | 1.18*** | 1.03*** | 0.76*** | 1.01*** |
| HML | 0.41*** | -0.82*** | -0.25 | 0.14 | -1.09* |
| SMB | -0.10 | -0.42*** | 0.16 | -0.38** | 0.12 |
| $f_0$ | 18694.95 | 5026.21 | 18226.81 | -61029.28 | 76400.12 |
| $f_1$ | 47819.56* | -63.27 | 5657.96 | 6895.08 | 14759.75 |
| $f_2$ | -34945.74 | -965.57 | 129450.70*** | -22110.56 | -19910.60 |
| $f_3$ | -11330.41 | | | | -64186.76** |
| $f_4$ | | | | | 158763.25* |
| $f_0^2$ | -5202.63 | -105.15 | -8498.80 | 45608.87 | 572.98 |
| $f_1^2$ | -13513.68** | -25.39 | 370.70 | 6895.08 | 1092.29 |
| $f_2^2$ | 3621.16 | 7.02 | -35508.48*** | -22110.56 | -1209.43 |
| $f_3^2$ | -11330.41 | | | | 4528.35*** |
| $f_4^2$ | | | | | -51473.61** |
| F | 1.03 | 0.73 | 2.08* | 0.35 | 2.77*** |

**Table 6.d.ii**

**e. French Fama with Buzzwords:**

$$E[R - R_f] = \alpha_i + \beta_1(R_M - R_f) + \beta_2 HML + \beta_3 SMB + \sum_{c=0}^{10} \beta_{w,c} w_c$$

**i. Using the Market as a benchmark:**

| | WMT | MSFT | NKE | PFE | TSLA |
|---|---|---|---|---|---|
| alpha | 0.01 | 0.10 | -0.08 | 0.10 | 1.04** |
| $w_0$ | 311.43 | -6321.19** | 214900.62*** | 395002.93* | -93918.89*** |
| $w_1$ | 82744.47 | 2624.92** | 92575.83*** | -583383.19*** | -40369.46 |
| $w_2$ | 56589.50*** | -5207.23 | 18722.72 | -27578.31 | 177504.46*** |
| $w_3$ | -106893.02 | 4901.61 | 12044.36 | -120562.47 | -32981.29 |
| $w_4$ | -53974.45*** | 5340.52 | -75339.64*** | -18672.24 | -21104.32 |
| $w_5$ | -69530.84 | -9066.98*** | 10284.35 | 134875.32 | -4932.53 |
| $w_6$ | 46702.14*** | 1553.39 | -139033.19** | 525891.78** | 279.21 |
| $w_7$ | -24933.27 | -6298.26* | 1338.70 | -43537.57 | 22459.06 |
| $w_8$ | -112041.50 | 10527.23*** | 217756.96* | -13777.55 | -1216.91 |
| $w_9$ | -24566.27 | 5849.92 | -357773.60** | -63065.44 | 82035.87** |
| Rm | 0.50*** | 1.04*** | 0.74*** | 0.62*** | 1.02*** |
| HML | 0.22* | -0.06 | -0.86*** | -0.05 | -0.39 |
| SMB | -0.44*** | -0.18 | 0.31 | -0.72*** | -0.00 |
| F | 5.76*** | 4.90*** | 7.14*** | 2.40** | 7.80*** |

**Table 6.e.i**

**ii. Using the ETF as a benchmark**

| | WMT | MSFT | NKE | PFE | TSLA |
|---|---|---|---|---|---|
| alpha | 0.04 | 0.05 | 0.02 | 0.14* | 1.00** |
| $w_0$ | -981.84 | -4665.50 | 213074.45*** | 306684.96 | -89372.13*** |
| $w_1$ | 113917.56* | 1370.06 | 77964.78*** | -547456.52*** | -52904.33* |
| $w_2$ | 43413.99** | -7353.58 | 43609.63 | -51596.42 | 167884.38*** |
| $w_3$ | -96596.38 | 5369.11 | 49839.43 | -60303.39 | -34453.43* |
| $w_4$ | -44432.29** | 1415.36 | -70377.92*** | 25978.42 | -23099.19 |
| $w_5$ | -77761.53 | -8767.53*** | -7641.90 | 141696.14 | 1057.70 |
| $w_6$ | 52419.96*** | 4461.49 | -167894.78*** | 531488.13** | 64.53 |
| $w_7$ | -32548.93 | -4520.97 | 44278.65 | -38175.85 | 14830.30 |
| $w_8$ | -233748.25* | 9128.96*** | 113620.70 | -28522.77 | 6158.75 |
| $w_9$ | -16874.62 | 4713.38 | -312241.69** | -24675.44 | 90143.22*** |
| Rm | 0.56*** | 1.24*** | 1.02*** | 0.71*** | 1.32*** |
| HML | 0.42*** | -0.76*** | -0.19 | 0.12 | -0.14 |
| SMB | -0.02 | -0.36** | 0.20 | -0.51*** | 0.00 |
| F | 5.35*** | 2.95*** | 8.61*** | 2.69*** | 7.53*** |

**Table 6.e.ii**

**f. French-Fama with sums of all comments lagged by one day:**

$$E[R - R_f] = \alpha_i + \beta_1(R_M - R_f) + \beta_2 HML + \beta_{i,3} SMB + \beta_{all} f_{all,t}$$

i. **Using the Market as a benchmark:**

| | WMT | MSFT | NKE | PFE | TSLA |
|---|---|---|---|---|---|
| alpha | 0.25** | -0.10 | -0.10 | 0.14 | 0.17 |
| Rm | 0.51*** | 1.04*** | 0.67*** | 0.68*** | 0.74*** |
| HML | 0.21 | -0.14 | -0.88*** | -0.05 | -1.09* |
| SMB | -0.55*** | -0.24* | 0.28 | -0.66*** | -0.21 |
| $f_{all}$ | -16441.29*** | 378.21** | 3812.63 | -19831.47 | 0.81 |
| F | 10.25*** | 4.71** | 0.84 | 1.09 | 0.00 |

**Table 6.f.i**

ii. **Using the ETF as a benchmark:**

| | WMT | MSFT | NKE | PFE | TSLA |
|---|---|---|---|---|---|
| alpha | 0.34*** | -0.25** | 0.00 | 0.17* | 0.16 |
| Rm | 0.61*** | 1.25*** | 1.04*** | 0.75*** | 0.85** |
| HML | 0.46*** | -0.84*** | -0.20 | 0.13 | -0.98 |
| SMB | -0.13 | -0.40*** | 0.19 | -0.41** | -0.21 |
| $f_{all}$ | -20369.98*** | 553.42*** | 4815.79 | -2275.32 | -214.93 |
| F | 14.67*** | 8.07*** | 1.72 | 0.01 | 0.00 |

**Table 6.f.ii**

g. **French-Fama with categorized comments lagged by one day:**

$$E[R - R_f] = \alpha_i + \beta_1(R_M - R_f) + \beta_2 HML + \beta_3 SMB + \sum_{c=0}^{k} \beta_{f,c} f_{c,t}$$

i. **Using the Market as a benchmark:**

| | WMT | MSFT | NKE | PFE | TSLA |
|---|---|---|---|---|---|
| alpha | 0.21* | -0.11 | -0.08 | 0.14 | 0.08 |
| Rm | 0.50*** | 1.01*** | 0.66*** | 0.68*** | 0.61** |
| HML | 0.21 | -0.19 | -0.88*** | -0.05 | -1.16** |
| SMB | -0.53*** | -0.27** | 0.29 | -0.66*** | -0.26 |
| $f_0$ | -2813.49 | 499.10 | -5249.26 | -18307.23 | 46157.60 |
| $f_1$ | -23145.89** | 3196.00*** | 12187.68 | -19359.85 | -6588.16 |
| $f_2$ | -23928.10 | -428.45 | -20092.23 | -27634.89 | -33763.11** |
| $f_3$ | 33509.06 | | | | 17263.69 |
| $f_4$ | | | | | 58650.06 |
| F | 3.31** | 3.53** | 0.50 | 0.36 | 1.51 |

**Table 6.g.i**

**ii. Using the ETF as a benchmark**

| | WMT | MSFT | NKE | PFE | TSLA |
|---|---|---|---|---|---|
| alpha | 0.31*** | -0.26** | 0.01 | 0.17* | 0.07 |
| Rm | 0.60*** | 1.20*** | 1.05*** | 0.75*** | 0.75* |
| HML | 0.46*** | -0.88*** | -0.22 | 0.13 | -1.05* |
| SMB | -0.11 | -0.45*** | 0.20 | -0.40** | -0.26 |
| $f_0$ | -7926.43 | -180.39 | -24461.04 | 3260.09 | 51846.03* |
| $f_1$ | -25866.57** | 4901.11*** | 17799.91 | -16552.65 | -8008.72 |
| $f_2$ | -26707.38 | -350.33 | -1136.31 | 2027.61 | -37160.21** |
| $f_3$ | 9234.08 | | | | 21692.70 |
| $f_4$ | | | | | 54079.55 |
| F | 4.00*** | 6.75*** | 0.98 | 0.06 | 1.81 |

**Table 6.g.ii**

**h. French-Fama with sums of all comments and squared sums lagged by one day:**

$$E[R-R_f] = \alpha_i+\beta_1(R_M-R_f)+\beta_2 HML+\beta_3 SMB+\beta_{fall} f_{all,t}+\beta_{all2} f^2_{all2,t}$$

**i. Using the Market as a benchmark:**

| | WMT | MSFT | NKE | PFE | TSLA |
|---|---|---|---|---|---|
| alpha | 0.16 | 0.29** | -0.17 | 0.14 | 1.64* |
| Rm | 0.51*** | 1.01*** | 0.67*** | 0.68*** | 0.66** |
| HML | 0.22* | -0.17 | -0.84*** | -0.05 | -1.07* |
| SMB | -0.55*** | -0.30** | 0.28 | -0.66*** | -0.18 |
| $f_{all}$ | 3634.56 | -1607.36*** | 15776.56 | -25601.42 | -19844.67** |
| $f^2_{all}$ | -3479.72* | 13.13*** | -477.48 | 2715.75 | 350.78** |
| F | 6.70*** | 9.88*** | 1.03 | 0.55 | 2.75* |

**Table 6.h.i**

**ii. Using the ETF as a benchmark:**

| | WMT | MSFT | NKE | PFE | TSLA |
|---|---|---|---|---|---|
| alpha | 0.27** | 0.28* | -0.07 | 0.17* | 1.86** |
| Rm | 0.61*** | 1.22*** | 1.04*** | 0.75*** | 0.85** |
| HML | 0.46*** | -0.83*** | -0.15 | 0.13 | -0.89 |
| SMB | -0.12 | -0.47*** | 0.18 | -0.41** | -0.18 |
| $f_{all}$ | -5776.71 | -2103.90*** | 16827.49 | -1884.50 | -22894.09** |
| $f_{all}^2$ | -2529.98 | 17.55*** | -479.19 | -184.53 | 400.41*** |
| F | 8.10*** | 15.98*** | 1.65 | 0.01 | 3.60** |

**Table 6.h.ii**

i. **French-Fama with categorized comments and comments squared lagged by one day:**

$$E[R - R_f] = \alpha_i + \beta_1(R_M - R_f) + \beta_2 HML + \beta_3 SMB + \sum_{c=0}^{k} \beta_{f,c} f_{c,t} + \sum_{c=0}^{k} \beta_{f,c2} f_{c2,t}$$

i. **Using the Market as a benchmark:**

| | WMT | MSFT | NKE | PFE | TSLA |
|---|---|---|---|---|---|
| alpha | 0.20* | 0.25 | -0.18 | 0.15* | 1.84** |
| Rm | 0.51*** | 1.00*** | 0.70*** | 0.69*** | 0.52** |
| HML | 0.22* | -0.17 | -0.80*** | -0.06 | -1.18** |
| SMB | -0.52*** | -0.29** | 0.30 | -0.67*** | -0.37 |
| $f_0$ | -22611.04 | 1402.67 | 9193.18 | -76179.05 | -139661.28** |
| $f_1$ | 3003.81 | -1630.43 | 26691.06 | -11784.12 | -59882.37** |
| $f_2$ | -7714.37 | -2366.84 | 28058.62 | -13172.64 | 15244.13 |
| $f_3$ | 14244.40 | | | | 15391.69 |
| $f_4$ | | | | | -109901.14 |
| $f_0^2$ | 8068.03 | -10.57 | 2132.98 | 34973.30 | 31599.25*** |
| $f_1^2$ | -7806.63 | 58.24* | -1407.21 | -11784.12 | 2682.49 |
| $f_2^2$ | -6700.96 | 27.37* | -19123.71 | -13172.64 | -2524.20* |
| $f_3^2$ | 14244.40 | | | | 915.02 |
| $f_4^2$ | | | | | 31101.55 |
| F | 2.31** | 3.25*** | 0.64 | 0.36 | 3.79*** |

**Table 6.i.i**

ii. **Using the ETF as a benchmark**

| | WMT | MSFT | NKE | PFE | TSLA |
|---|---|---|---|---|---|
| alpha | 0.29** | 0.16 | -0.07 | 0.18* | 1.98** |
| Rm | 0.60*** | 1.20*** | 1.05*** | 0.76*** | 0.56 |
| HML | 0.48*** | -0.85*** | -0.15 | 0.13 | -1.14** |
| SMB | -0.10 | -0.48*** | 0.20 | -0.41** | -0.40 |
| $f_0$ | -26690.46 | -3253.09 | -27246.71 | -45341.36 | -135517.20** |
| $f_1$ | 7220.64 | -291.43 | 36982.56 | -10039.26 | -63323.64** |
| $f_2$ | -24951.35 | -1228.22 | 29748.78 | 1664.90 | 4709.49 |
| $f_3$ | 1710.76 | | | | 21470.69 |
| $f_4$ | | | | | -111652.22 |
| $f_0^2$ | 7466.28 | 69.14 | 6594.89 | 29468.07 | 31783.76*** |
| $f_1^2$ | -9375.59 | 60.68 | -2102.31 | -10039.26 | 2839.95 |
| $f_2^2$ | -2046.06 | 21.18 | -13158.61 | 1664.90 | -2067.34 |
| $f_3^2$ | 1710.76 | | | | 736.28 |
| $f_4^2$ | | | | | 30240.14 |
| F | 2.70** | 5.09*** | 0.84 | 0.10 | 3.91*** |

**j. French Fama with Buzzwords lagged by one day:**

$$E[R - R_f] = \alpha_i + \beta_1(R_M - R_f) + \beta_2 HML + \beta_3 SMB + \sum_{c=0}^{10} \beta_{w,c} w_c$$

**i. Using the Market as a benchmark:**

| | WMT | MSFT | NKE | PFE | TSLA |
|---|---|---|---|---|---|
| alpha | 0.19* | -0.09 | -0.14 | 0.14* | 0.25 |
| $w_0$ | -3624.40 | 2478.11 | -34896.24 | 231190.37 | -27362.24 |
| $w_1$ | -22083.45 | 1718.65 | -81960.22** | -76499.44 | -8336.45 |
| $w_2$ | -9131.80 | -1283.84 | 188246.32*** | -86033.44 | 36844.45 |
| $w_3$ | -27680.27 | -1223.01 | 67203.54 | 135054.21 | -40597.75 |
| $w_4$ | -4166.69 | -6562.64 | 13220.39 | -148964.49** | -4603.93 |
| $w_5$ | -104439.66 | -1656.78 | 111489.54*** | -117960.26 | -56554.89 |
| $w_6$ | -17105.48 | 3315.66 | 149533.45** | 48620.97 | 1008.59 |
| $w_7$ | -23181.63 | 2440.03 | -84199.41 | 51270.22 | 27526.45 |
| $w_8$ | -113733.47 | -2019.52 | -420377.62*** | -30792.53 | 29870.81 |
| $w_9$ | -13581.22 | -1531.58 | -81078.85 | -106347.75 | 24881.31 |
| Rm | 0.49*** | 1.03*** | 0.69*** | 0.66*** | 0.73*** |
| HML | 0.21 | -0.13 | -0.74*** | -0.09 | -1.07* |
| SMB | -0.48*** | -0.25* | 0.51* | -0.65*** | 0.05 |
| F | 1.64 | 1.30 | 4.11*** | 1.20 | 1.26 |

**Table 6.j.i**

#### ii. Using the ETF as a benchmark

| | WMT | MSFT | NKE | PFE | TSLA |
|---|---|---|---|---|---|
| alpha | 0.25** | -0.27** | -0.01 | 0.17* | 0.25 |
| $w_0$ | 21184.74 | 3461.16 | -36616.92 | 30206.92 | -32774.52 |
| $w_1$ | 1858.98 | 4472.14*** | -88311.18*** | 66394.76 | -12193.55 |
| $w_2$ | -25046.72 | -1326.37 | 199077.30*** | -13171.13 | 43594.16 |
| $w_3$ | -104839.29 | -473.81 | 74211.76 | 224246.78 | -42937.11* |
| $w_4$ | -8773.22 | -6739.31 | 3542.13 | -137490.43* | -7002.51 |
| $w_5$ | -208968.02** | -3997.08 | 97871.25*** | -179388.39* | -54273.79 |
| $w_6$ | -10429.06 | 4072.11 | 175148.13*** | -113435.58 | -1156.81 |
| $w_7$ | -9430.06 | -3695.26 | -109279.65* | 125109.33* | 33017.96 |
| $w_8$ | -166253.44 | -3264.12 | -363038.63*** | -27264.66 | 31523.85 |
| $w_9$ | -32796.45 | -4033.24 | -40000.75 | 305.81 | 25467.86 |
| Rm | 0.58*** | 1.23*** | 1.05*** | 0.73*** | 0.96** |
| HML | 0.46*** | -0.86*** | -0.08 | 0.10 | -0.87 |
| SMB | -0.04 | -0.42*** | 0.41* | -0.41** | 0.09 |
| F | 2.58*** | 2.36** | 6.05*** | 1.16 | 1.41 |

**Table 6.j.ii**

## 7. Discussion of Results

Use of an ETF as opposed to the overall market benchmark didn't often change the significance of any results, which is not surprising, given that industry ETFs are generally correlated with the overall market (The IYW technology ETF had a correlation coefficient with the market of 0.94, which was the highest, while the lowest was the VDC consumer goods ETF, which had a correlation coefficient of 0.65). There were a few cases where changing the benchmark changed the impact of other parameters. For example, comparing Table 6.e.i to Table 6.e.ii, it's clear that the impact of w_0 and w_1 is not statistically significant once the ETF is used as a benchmark. In these few cases, where using an industry-specific benchmark makes specific comments irrelevant, I theorize that comments are providing an intermediate amount of information: they reflect or are in reaction to market movements that affect a stock and its competitors.

Notably, for both benchmarks, for all stocks, models using buzzwords were significant. The impact of these factors is neither enormous, nor trivial in magnitude. For example, in table Table 6.e.i, in the row corresponding to w_6, in the column corresponding to WMT, the beta coefficient is 46702.14. For Walmart, w_6 is the word "puts" This means that on a day with 150,000 comments on /r/wallstreetbets (the average during the period from June 1st, 2018 to December 10, 2018 was 115826.79), for every 5 times the word “puts” appears in a comment that mentions Walmart or WMT, we expect the daily returns of Walmart to increase by 155 basis points.

In some cases, reddit factors were not significant unless lagged by one day. For example, comparing Table 6.b.i to Table 6.g.i, classified comments don’t impact Microsoft stock on the day of, but the comments from the previous day evidently do. The behavior that these comments are either anticipating or provoking may take some time to manifest. The reddit effect is not always instantaneous.

Tesla stands out as the most susceptible to Reddit factors. It was the only stock which classified comments and daily word counts were useful for anticipating daily returns. Classified comments were still significant even after lagging them by one day, showing their effect had some persistence. This is consistent with recent analysis by Barclays that the price of Tesla stock was being inflated by /r/wallstreetbets. (Mckay 2021). As was stated earlier, the fundamentals of Tesla are hard to gauge given its involvement in many industries, its price is heavily weighted on future earnings, and the bombastic CEO, Elon Musk (who is by some considered a genius, and by others considered a sociopath). Some of Musk’s tweets have gotten him in trouble for fraud with the SEC, showing that there is legitimate concern that the things people read on social media will direct their trading behavior. It is befitting that these social media rallies could be driven by the sophomoric, uncouth denizens of reddit as well as the equally sophomoric, uncouth CEO of the stocks affected.

## 8. Conclusion

Ultimately, the inclusion of /r/wallstreetbets factors into a Fama-French model for predicting daily returns of stocks enhances the model. It is ambiguous as to whether this can be seen as /r/wallstreetbets egging each other on to buy or sell a particular stock en masse, or sophisticated investors discussing the fundamental and technical factors that cause market movements that don’t originate on reddit. Both could coexist at any given moment. The same messages can be seen as trying to incite a herd as well as fundamental analysis. One prominent Youtuber and reddit user, named Keith Gill is now being sued for fraud (McKay, 2021). He had been very bullish on GameStop, claiming it was undervalued due to its good leadership and its position in the massive gaming industry. In this way, Gill was simply a standard value investor like Warren Buffett, finding and investing in a “wonderful company at a good price”. Nonetheless, the suit against Gill claims he caused “enormous losses not only to those who bought option contracts, but also to those who fell for Gill's act and bought GameStop stock during the market frenzy at greatly inflated prices" (McKay, 2021).

There are many ways this project could be expanded upon. I could use different techniques using current data. There are more sophisticated ways to represent documents than a bag-of-words, such by using a Character to Sentence Convolutional Neural Network for feature extraction (Dos Santos & Gotti 2014), which embeds information about characters in the words they make up, and embeds information about words in the sentences they compromise. Reducing dimensionality can be done by Autoencoders, a type of Neural Network. I could also use a different unsupervised clustering method, like Deep Embedding Clustering, (Xie Girshick & Farhadi 2016), which simultaneously optimizes the parameters of an autoencoder and cluster centers to minimize distance between the reduced document vectors and the centers of their cluster.

I could widen the scope and look at a wider basket of securities. Other stocks that fit the criteria of being well-known, and not sharing the name of any other common words include Nvidia, Goldman Sachs and Netflix. I could even circumvent the homonymy problem, thus allowing me to analyze corporations with names like Apple, Alphabet and Visa, by training a classifier to label text that contains the names in question as discussing the corporation or otherwise. This would be a supervised classification problem, with the labeled training data chosen heuristically. For example, to train the Apple classifier, I would retrieve financial news articles that refer to the Apple corporation, and then text samples from books about cooking or botany that refer to the fruit, and manually label each document. With this dataset, I could use any model that can do binary classification like Perceptron, Support Vector Machine or a Neural Network. Once the model is trained, it could classify all of the reddit comments as discussing the company in question or not, and discard the comments that are classified as not relevant.